\documentclass[12pt,a4paper]{article}

\usepackage{latexsym}
\usepackage{amsmath}
\usepackage{amsfonts}
\usepackage{amssymb}
\usepackage{amsthm}
\usepackage{amscd}
\usepackage{mathrsfs}

\newcommand{\be}{\begin{equation}}
\newcommand{\ee}{\end{equation}}
\newcommand{\bea}{\begin{eqnarray}}
\newcommand{\eea}{\end{eqnarray}}

\renewcommand{\theequation}{\arabic{section}.\arabic{equation}}

\def\cJ{{\cal J}}                       %
\def\cR{{\cal R}}                       %
\def\cG{{\mathfrak{g}}}                 %
\def\cK{{\mathfrak{K}}}                 %
\def\cH{{\mathcal H}}                   %
\def\red{\mathrm{red}}                  %
\def\bR{{\mathbb R}}                    %
\def\tr{\mathrm{tr}}                    %
\def\diag{\mathrm{diag}}                %
\begin{document}

\vspace*{0.5cm}

\begin{center}
{\Large \bf Action-angle map and duality for the open Toda lattice
in the perspective of Hamiltonian reduction}
\end{center}

\vspace{0.2cm}

\begin{center}
L. Feh\'er
  \\

\bigskip

Department of Theoretical Physics, University of Szeged\\
Tisza Lajos krt 84-86, H-6720 Szeged, Hungary, and\\
Department of Theoretical Physics, WIGNER RCP, RMKI \\
H-1525 Budapest, P.O.B.~49,  Hungary\\
 e-mail: lfeher@physx.u-szeged.hu

\bigskip

\bigskip

\end{center}

\vspace{0.2cm}

\begin{abstract}
An alternative derivation of the known action-angle map of the standard open Toda lattice
is presented based on its identification as the natural map between two gauge slices in
the relevant symplectic reduction of the cotangent bundle of $GL(n,\bR)$.
This then permits to interpret Ruijsenaars' action-angle duality for the Toda system in the
same group-theoretic framework which  was established previously for Calogero type systems.
\end{abstract}


\newpage
\section{Introduction}

It is common knowledge that Liouville integrable systems admit action-angle variables which
trivialize the flows of the commuting Hamiltonians.
There exist powerful methods (see e.g.~\cite{BBT})  to construct action-angle variables
from the original variables, but then one still has to face the
difficult and relevant problem of reconstructing the original variables from the action-angle variables.
For purely scattering systems,  one can in principle use
the asymptotic momenta and their conjugates to  obtain globally well-defined action-angle variables.
However, even among  such topologically trivial systems
 it is very exceptional that one can
describe the map from the action-angle variables to the original variables
\emph{explicitly}.

A beautiful example
is provided by the standard open Toda lattice encoded
as a Hamiltonian system by $(M,\omega,H)$, where
$ M:= \bR^n \times \bR^n$ with the Darboux form
$\omega = \sum_{i=1}^n d p_i \wedge d q_i$ and
\be
H(q,p)=\frac{1}{2}\sum_{i=1}^n p_i^2 + \sum_{i=1}^{n-1} e^{q_i - q_{i+1}}.
\label{1.1}\ee
The phase space of the corresponding action-angle variables,
here
denoted by
$\hat p$ and $\hat q$, is
\be
\hat M := \{ (\hat p, \hat q) \in \bR^n \times \bR^n \,\vert\, \hat p_1 > \hat p_2 > \cdots > \hat p_n\}
\label{1.2}\ee
equipped with the symplectic form
$\hat \omega = \sum_{i=1}^n d \hat q_i \wedge d \hat p_i$.
One can find the \emph{explicit} action-angle map, $R\colon \hat M \to M$, in the paper
\cite{SR-90} by Ruijsenaars\footnote{Our variables $(\hat p, \hat q)$ correspond to
$(-\hat \theta, \hat q)$ as used in \cite{SR-90}.}.
The map $R$ operates according to the following formula:
\be
q_j = \ln(\sigma_{n+1-j}/\sigma_{n-j}),
\qquad
p_j= \dot{\sigma}_{n+1-j}/\sigma_{n+1-j}- \dot{\sigma}_{n-j}/\sigma_{n-j},
\quad \forall j=1,\ldots, n,
\label{1.3}\ee
where $\sigma_0:=1$ and
\be
\sigma_k:= \sum_{\vert I \vert =k} e^{\sum_{l\in I} \hat q_l} \!\! \prod_{i\in I, j\notin I}
\vert \hat p_i - \hat p_j\vert^{-1},
\qquad \forall k=1,\ldots, n.
\label{1.4}\ee
The sum is taken over the subsets $I\subset \{1,2,\ldots,n\}$ of cardinality $\vert I\vert =k$
and $\dot \sigma_k$ is defined by
$\dot \sigma_k := \{\sigma_k, \frac{1}{2} \sum_{i=1}^n \hat p_i^2\}_{\hat M}$.
This map $R$ converts $H$ into the free form
$H\circ R = \frac{1}{2} \sum_{i=1}^n \hat p_i^2$.
Ruijsenaars' derivation \cite{SR-90} relied on scattering theory and careful analysis
of the Toda dynamics both for obtaining the map $R$ and for proving that it is
a symplectomorphism.
His construction was inspired by
the pioneering work of Moser \cite{Mos}  and formulas from
the papers of Kostant \cite{Kos} and Olshanetsky-Perelomov \cite{OP} dealing with the explicit
solution of  the open Toda systems attached to all simple root systems \cite{Bog}.
For reviews of Toda systems, see e.g.~\cite{BBT,Per,DLT,RSTS}.

The first goal of this Letter is to present an alternative derivation
of the Toda action-angle map.
We recall \cite{OP} (or \cite{BBT,Per,RSTS})  that the system (\ref{1.1}) results from
Hamiltonian
reduction of the free geodesic motion
 on the symmetric space $GL(n,\bR)/O(n)$, and  thus it can be also viewed as a reduction
 of free motion on the larger configuration space $GL(n,\bR)$.
We shall explain how this basic fact leads readily to the map $R$ and all its
pertinent properties.

The second (and main)  goal is to demonstrate  that the action-angle dual
 of the Toda system,  introduced in \cite{SR-90}, can be interpreted in the same
 manner as is now well-known (see e.g.~\cite{JHEP,FK-NPB}) regarding
Ruijsenaars' duality relations \cite{SR-88,RIMS95} between Calogero type systems.
 Recall that two integrable many-body systems form a dual pair  if their phase spaces
are related by a symplectomorphism that converts the particle-positions and the action-variables
of one system, respectively, into the action-variables and the particle-positions of the other system.
The usual group-theoretic picture behind this kind of duality
(called action-angle duality or Ruijsenaars duality)
is as follows.
There exists a ``big phase space'' equipped with two distinguished Abelian
Poisson algebras which descend to particle-positions and action-variables of
integrable many-body systems upon a suitable reduction.
The key point is that the single
 reduced phase space admits two alternative models, typically given by
two gauge slices, which are identified with the phase spaces of the
many-body systems in duality.
In particular,
the roles of the two reduced Abelian Poisson algebras
as positions and actions are interchanged in the two models.

In fact, we shall identify the phase spaces $(M,\omega)$ and $(\hat M, \hat \omega)$
with two gauge slices  in
the relevant symplectic reduction of $T^* GL(n,\bR)$.
The Toda system lives on $M$, while $\hat M$ supports the dual many-body Hamiltonian
defined by
\be
\hat H(\hat p, \hat q) := \sigma_1(\hat p,\hat q) =
\sum_{i=1}^n e^{\hat q_i}
\prod^n_{\substack{j=1\\j \neq i}}
\frac{1}{\vert \hat p_i - \hat p_j\vert}.
\label{1.5}\ee
We shall then explain that the geometrically engendered symplectomorphism between the two
slices yields the action-angle map $R$, and confirm that the above sketched
interpretation of the duality relation holds in the Toda case.
Conceptually, the whole picture is the same as for Calogero type systems.
The special feature is that now the map
from $\hat M$ to $M$ can be described fully explicitly.

The description of the reduction picture of the Toda duality is the principal achievement
of this Letter.
This represents a step forward in the research program
aimed at understanding all action-angle dualities in group-theoretic terms based
on Hamiltonian reduction.
As for the action-angle map itself,  it must be stressed that several other alternative derivations
are possible.
For example, derivations of the  Toda action-angle variables
can be found in \cite{DLNT,FG,Tsig,Skly},
although the  author is not aware of any reference other than \cite{SR-90} where the
reconstruction of the original variables is given
in as explicit form as displayed in equations (\ref{1.3}), (\ref{1.4}) above.

The rest of this text is organized as follows. In subsections 2.1 and 2.2,  we  present two models of
the reduced phase space: one is the
standard Toda slice \cite{OP,Per} and the other is a new one, which we shall
term ``Moser gauge".
In subsection 2.3, we derive the explicit action-angle  symplectomorphism $R$ as the map
from the Moser gauge to the Toda gauge.
In section 3,
we explain the duality issues and conclude with
 comments  on open problems.

\section{Two descriptions of the reduced phase space}
\setcounter{equation}{0}

In what follows we denote $G:=GL(n,\bR)$ and let
$\cG$ stand for the corresponding Lie algebra $gl(n,\bR)$.
We consider the maximal compact subgroup
$K:=O(n,\bR)$, the group $A$ of positive diagonal matrices and the group
$N_+$ of upper  triangular matrices having 1 along the diagonal.

We are going to  reduce the cotangent bundle of $G$,
which we realize as
\be
T^* G \simeq G \times \cG = \{ (g, \cJ)\},
\label{2.1}\ee
where left-translations are used for trivializing $T^* G$ and    $\cG^*$ is identified with $\cG$
by means of the invariant bilinear form of $\cG$ provided by the matrix trace.
 The Hamiltonian of the free particle moving on $G$ is the $k=2$ member of the Poisson commuting family
\be
\cH_k(g,\cJ):= \frac{1}{k}\tr(\cJ^k),
\qquad
k=1,\ldots, n.
\label{cH2}\ee
The symmetry group whereby we reduce is the direct product
$N_+ \times K$.
An arbitrary element $(\eta_+, \eta_K)$ from this group acts on $T^*G$
by the map $\Psi_{(\eta_+, \eta_K)}$ defined by
\be
\Psi_{(\eta_+,\eta_K)}(g,\cJ) := (\eta_+ g \eta_K^{-1}, \eta_K \cJ \eta_K^{-1}).
\label{2.4}\ee
We equip $T^*G$ with the symplectic form
\be
\Omega :=2 d \tr (\cJ g^{-1} dg),
\label{2.5}\ee
which is invariant under $\Psi_{(\eta_+,\eta_K)}$.
The action of $N_+ \times K$ is generated by a moment map, $\Phi$.
To describe the moment map,
we use the vector space decompositions
\be
\cG = \cG_+ + \cG_0 + \cG_-
\quad\hbox{and}\quad
\cG = \cK  + \cK^\perp,
\label{2.6}\ee
where the first one refers to strictly upper-triangular, diagonal and lower-triangular matrices,
while $\cK$ and $\cK^\perp$ consist of antisymmetric and symmetric matrices, respectively.
Accordingly,  we can decompose any $X\in \cG$ as
\be
X= X_+ + X_0 + X_-
\quad\hbox{and}\quad
X = X_\cK + X_{\cK^\perp}.
\label{2.7}\ee
By using the trace scalar product, the duals of the Lie algebras of $N_+$ and $K$ can be identified with
$\cG_-$ and with $\cK$ itself.
Applying these conventions,
the moment map
\be
\Phi\colon T^*G \to (\cG_+)^* \times \cK^*\simeq \cG_- \times \cK
\label{2.8}\ee
operates as
\be
\Phi(g,\cJ) = ((g \cJ g^{-1})_-, -\cJ_\cK).
\label{2.9}\ee
The relevant moment map constraint then reads
\be
\Phi(g,\cJ) =\mu_0:= (I_-,0),
\label{2.10}\ee
where the matrix $I_-:= \sum_{i=1}^{n-1} E_{i+1,i}$ contains $1$ in its entries just below the diagonal.
It is easily seen that
the constraint surface $\Phi^{-1}(\mu_0)$
is preserved by the full symmetry group $N_+ \times K$.
In other words,  the constraints
\be
(g \cJ g^{-1})_- - I_-=0
\quad\hbox{and}\quad
\cJ_\cK=0
\label{constraints}\ee
are of first class in Dirac's sense.
The corresponding reduced phase space,
\be
(T^* G)_\red := \Phi^{-1}(\mu_0)/(N_+ \times K),
\label{2.12}\ee
is guaranteed to be a smooth manifold since the action of $N_+ \times K$ is free and proper.
This follows from the Iwasawa decomposition whereby any
$g\in G$ can be uniquely represented in the form
\be
g= g_+ g_A g_K,
\qquad (g_+, g_A, g_K) \in N_+ \times A \times K.
\label{2.13}\ee

\subsection{Toda gauge}

Taking arbitrary $(q,p) \in \bR^n \times \bR^n$, we now parametrize
$g_A$ in (\ref{2.13}) as
\be
g_A= e^{Q(q)/2}
\quad\hbox{with}\quad
Q(q): =  -\sum_{i=1}^n q_{n+1-i} E_{i,i},
\label{par}\ee
and introduce the Jacobi matrix $L(q,p)$ by
\be
L(q,p):= P(p) + e^{-Q(q)/2} I_-  e^{Q(q)/2}  + e^{Q(q)/2} I_+  e^{-Q(q)/2}
\label{Lqp}\ee
with $P(p):=- \sum_{i=1}^n p_{n+1-i} E_{i,i}$ and $I_+ := (I_-)^t$.
(The usage of $Q(q)$ and $P(p)$ is justified by later convenience.)
The Iwasawa decomposition implies that every orbit of
$N_+ \times K$ in $T^*G$ contains a unique representative
of the form
$(e^{Q(q)/2}, \cJ)$.
If $g=e^{Q(q)/2}$, then the constraint (\ref{constraints})  is solved
by  $\cJ= L(q,p)$.
This means that the manifold
\be
S:= \{ (e^{Q(q)/2}, L(q,p))\,\vert\, (q,p)\in M\},
\qquad M = \bR^n \times \bR^n,
\label{Sdef}\ee
is a \emph{global cross-section} (``gauge slice'') of the orbits of the symmetry group in
the constraint surface $\Phi^{-1}(\mu_0)$.
If we identify the reduced phase space $(T^*G)_\red$ with the gauge slice $S$, then
the reduced symplectic form, $\Omega_\red$,  turns into  the pull-back  $\iota_S ^* (\Omega)$,
where $\iota_S: S \to T^*G$ is the tautological inclusion.
One immediately finds that
\be
\iota_S ^* (\Omega) = \sum_{i=1}^n dp_i \wedge dq_i \equiv \omega,
\label{2.17}\ee
which is the Darboux form on $M$.
These observations are summarized by the  identifications
\be
((T^*G)_\red, \Omega_\red) \simeq (S, \iota_S^*(\Omega)) \simeq (M, \omega).
\label{2.18}\ee
The  equality
\be
H(q,p) = \frac{1}{2} \tr( L(q,p)^2)
\label{2.19}\ee
shows that the Toda Hamiltonian (\ref{1.1}) is the reduction of the free one, $\cH_2$ in (\ref{cH2}).
The matrix $L(q,p)$ appearing in the slice $S$ (\ref{Sdef})
is the standard Toda Lax matrix, and
for this reason  $S$ can be called the ``Toda gauge''.
Of course,  all the above  are very standard results \cite{Per,RSTS}.

\subsection{Moser gauge}

Let $\bR^n_>$ denote the set of vectors $\hat p \in \bR^n$ subject to the condition
$\hat p_1 > \hat p_2 >\cdots > \hat p_n$,
and  $\bR^n_+$ the set of vectors with positive components.
For any $(\hat p, w)\in \bR^n_> \times \bR^n_+$ define the matrices
\be
\Lambda(\hat p) := \diag(\hat p_1, \hat p_2,\ldots, \hat p_n),
\label{2.21}\ee
\be
\Gamma(\hat p, w) := [w, \Lambda(\hat p)w, \Lambda(\hat p)^2 w,\ldots, \Lambda(\hat p)^{n-1} w]
= W V(\hat p).
\label{2.22}\ee
Here $w$ is represented as a column vector, $W$ and $V$
are invertible diagonal and Vandermonde matrices
\be
W:= \diag(w_1,\ldots, w_n)
\quad\hbox{and} \quad V(\hat p)_{i,j} := (\hat p_i)^{j-1}.
\label{VW}\ee

The Toda gauge was defined by bringing
 $g\in G$ to diagonal form by the action of $N_+ \times K$.
Now we introduce another gauge, in which the component $\cJ$ of the pair $(g,\cJ)\in \Phi^{-1}(\mu_0)$ is
 diagonalized. In fact, we claim that
\emph{the manifold
\be
\hat S:= \{ (\Gamma(\hat p, w)^{-1}, \Lambda(\hat p))\,\vert\, (\hat p, w) \in \bR^n_> \times \bR^n_+\}
\label{hatSdef}\ee
is a global cross-section of the orbits of $N_+ \times K$ in the constraint surface $\Phi^{-1}(\mu_0)$.}

Note first that $\hat S$ indeed lies in
the constraint surface $\Phi^{-1}(\mu_0)$. This can be seen, for example, from the identity
 $\Lambda \Gamma - \Gamma I_- = [0, \ldots, 0, \Lambda^n w]$,
 which implies that
 $(\Gamma^{-1} \Lambda\Gamma)_-=I_-$ holds.
To proceed, consider
 the Iwasawa decomposition
\be
\Gamma(\hat p, w)^{-1} = \eta_+(\hat p, w) \rho(\hat p, w) \eta_K(\hat p, w)
\label{Iwas}\ee
with unique matrices $\eta_+(\hat p,w)\in N_+$, $\eta_K(\hat p,w) \in K$ and diagonal positive matrix
\be
\rho(\hat p, w) =  \diag(\rho_1(\hat p, w), \ldots, \rho_n(\hat p, w)).
\label{2.25}\ee
The factors of $\Gamma^{-1}$ enjoy the scaling properties
\be
\rho(\hat p, \lambda w) = \lambda^{-1} \rho(\hat p, w),
\quad
\eta_+(\hat p, \lambda w) = \eta_+(\hat p, w),
\quad
\eta_K(\hat p, \lambda w) = \eta_K(\hat p, w),
\quad \forall \lambda \in \bR_+.
\label{scale}\ee
Acting by $(\eta_+(\hat p,w)^{-1}, \eta_K(\hat p,w))\in N_+ \times K$,
we obtain
\be
\Psi_{ (\eta_+(\hat p,w)^{-1}, \eta_K(\hat p,w))}( \Gamma(\hat p, w)^{-1}, \Lambda(\hat p))=
(\rho(\hat p,w), \cJ(\hat p,w)),
\label{gt}\ee
\be
\cJ(\hat p,w)= \eta_K(\hat p, w) \Lambda(\hat p) \eta_K(\hat p,w)^{-1}.
\label{2.28}\ee
Then $(\rho(\hat p,w), \cJ(\hat p,w))$ belongs to the Toda gauge slice $S$ (\ref{Sdef}),
because $\rho(\hat p, w) \in A$ and the moment map constraint (\ref{constraints}) holds.
Therefore we have the equalities
\be
\cJ(\hat p,w) = L(q,p)
\quad\hbox{and}\quad
\rho(\hat p,w) = e^{Q(q)/2}
\label{2.29}\ee
with uniquely determined $(q,p) \in M$. Now we quote (see e.g.~\cite{DLT,Nanda})
the following well-known result:
\emph{every Jacobi matrix
$L(q,p)$ can be written in the form (\ref{2.28}) and this yields a one-to-one parametrization
of the set of Jacobi matrices after one fixes the norm of the vector $w$.}
Combining this result with the scaling properties (\ref{scale}), we see that
every element of the Toda gauge $S$ is obtained as a gauge transform (\ref{gt}) of a unique
element of $\hat S$.
Since we know that $S$ is a global cross-section, the claim
that $\hat S$ is global cross-section follows.

We call $\hat S$  ``Moser gauge'' since
the variables $(\hat p, w)$ were first introduced
in Moser's seminal work \cite{Mos} on the Toda system.
The reduced symplectic form is easily
calculated\footnote{The calculation, which the author learned
from C. Klim\v c\'\i k, is described in Appendix A for convenience. }
in Moser's variables.
One finds that
\be
\iota_{\hat S}^*( \Omega) = 2 \sum_{i=1}^n d\ln w_i \wedge  d \hat p_i +
 \sum^n_{\substack{j, k=1\\j\neq k}}\frac{d \hat p_j \wedge d\hat p_k}{\hat p_j - \hat p_k}.
\label{symp1}\ee
By viewing $\hat S$ as a coordinate system on $S$, which is
allowed since both are models of  $(T^*G)_\red$, the Toda Hamiltonian becomes
\be
(\iota_{\hat S}^* \cH_2)(\hat p, w) = \frac{1}{2} \sum_{i=1}^n \hat p_i^2.
\label{HR}\ee
Moser's variables linearize
the Toda dynamics, but are not quite action-angle variables since
their Poisson brackets read
\be
\{\hat p_i, \hat p_j \} =0,
\qquad
\{ \hat p_i, w_j\} = \frac{w_j}{2} \delta_{ij},
\qquad
\{ w_j, w_k\} =\frac{1}{2}\frac{w_j w_k}{\hat p_j - \hat p_k}\quad (j\neq k).
\label{2.32}\ee
One can construct true action-angle variables $\hat p\,$-$\,\hat q$ by using a primitive of the de Rham exact
 second  term
that appears in (\ref{symp1}).
This is of course not unique, and the choice which will be useful for our purpose
corresponds to the following parametrization:
\be
w_i(\hat p, \hat q): = e^{\frac{1}{2}\hat q_i} \prod^n_{\substack{j=1\\j \neq i}} \vert \hat p_i -
\hat p_j\vert^{-\frac{1}{2}},
\quad
(\hat p, \hat q)\in \bR^n_> \times \bR^n \equiv \hat M.
\label{wpar}\ee
Direct substitution shows that
\be
\iota_{\hat S}^*( \Omega) = 2\sum_{i=1}^n d\ln w_i(\hat p,\hat q) \wedge  d \hat p_i +
 \sum^n_{\substack{j,k=1\\j\neq k}}\frac{d \hat p_j \wedge d\hat p_k}{\hat p_j - \hat p_k}
= \sum_{i=1}^n d \hat q_i \wedge d\hat p_i \equiv \hat \omega.
\label{2.34}\ee
The identification
$(\hat S, \iota_{\hat S}^* (\Omega)) \equiv (\hat M, \hat \omega)$
given by the above equations
anticipates that $(\hat p, \hat q)$ as defined here
coincide with the action-angle variables of \cite{SR-90} described in the Introduction.

For completeness, we note that Moser's variables are usually presented by means of the resolvent function
\be
f(z):=  e^{q_n} ((z - L(q,p))^{-1})_{1,1} = \sum_{i=1}^n \frac{w_i^2}{z- \hat p_i},
\label{res}\ee
where $z$ is an auxiliary complex variable.
The second equality, which originally served as the definition of $(\hat p,w)$,
 follows from (\ref{2.29}).
To see how this comes about,  notice that equations (\ref{2.28}) and (\ref{2.29}) imply the
 relation $((z - L(q,p))^{-1})_{1,1} = \sum_{i=1}^n \frac{((\eta_K)_{1,i})^2}{z- \hat p_i}$.
 Then substitute the relation $(\eta_K)_{1,i} = w_i \rho_1 = w_i e^{-q_n/2}$, which
 is obtained from the Iwasawa decomposition of $\Gamma$ (cf.~(\ref{Iwas}))
 and  the second equality in (\ref{2.29}).
 By using the identity (\ref{res}),  one can verify that the variables $(\hat p, \hat q)$
coincide also with the action-angle variables exhibited in \cite{FG,Tsig} relying on
an interesting recasting of
the Toda Poisson brackets in terms of the resolvent function.

\subsection{Explicit action-angle map from gauge transformation}

The results described so far give rise to a symplectomorphism
between the two models,
\be
(S, \iota_S^*(\Omega))\equiv (M,\omega)
\quad\hbox{and}\quad
(\hat S, \iota_{\hat S}^*(\Omega)) \equiv (\hat M, \hat \omega),
\label{ident}\ee
of the reduced phase space $((T^*G)_\red, \Omega_\red)$.
Under this symplectomorphism, any point of $\hat S$ corresponds to the unique
gauge equivalent point of $S$.
It turns out that, when
operating in the direction $\hat S \to S$, the explicit formula of this map
is easily computed.
We below denote the map in question as $\cR\colon \hat S \to S$, and will
in the end identify it with  the action-angle map $R\colon \hat M \to M$
presented in the Introduction.

Let $m_k(X):= \det(X_k)$ denote the $k$-th leading principal minor of any $n \times n$ matrix $X$, i.e.,
the determinant of the matrix $X_k$ obtained by deleting the last $(n-k)$ rows and columns of $X$.
Introduce the following Poisson commuting Hamiltonians on $T^*G$:
\be
\hat \cH_k(g,\cJ):= m_k( (g g^t)^{-1}),
\qquad
k=1,2,\ldots, n.
\label{hatcHk}\ee
These Hamiltonians are invariant with respect to the
action of the reduction group $N_+ \times K$.

If $\cR$ sends a point $( \Gamma^{-1}, \Lambda)\in \hat S$ to  $(e^{Q(q)/2}, L(q,p))\in S$, then
the equality
\be
m_k (\Gamma^t \Gamma)  = \prod_{j=1}^k e^{q_{n+1-j}}
\label{mk}\ee
holds, since on the two sides we have the values of the invariant
function $\hat \cH_k$ at gauge equivalent points.
By (\ref{ident}), the components of $q$ and $p$, and that of $\hat p$ and $\hat q$,
give coordinates on $S$ and on $\hat S$, respectively.
We then see from (\ref{mk}) that the
sought-after formula of $\cR$ is provided by
\be
e^{q_k}\circ \cR =
\frac{ m_{n+1-k}(X(\hat p,\hat q))}{m_{n-k}(X(\hat p, \hat q))}
\quad
\hbox{with}\quad
X(\hat p, \hat q):= \Gamma(\hat p, w(\hat p, \hat q))^t \Gamma(\hat p, w(\hat p, \hat q)),
\label{qkcircR}\ee
\be
p_k\circ \cR  = \{q_k, H\}_M \circ \cR = \{ q_k \circ \cR, H \circ \cR \}_{\hat M}
= \{ q_k \circ \cR, \frac{1}{2} \sum_{i=1}^n \hat p_i^2\}_{\hat M}.
\label{pcircR}\ee
We here used that $\cR$ is a symplectomorphism
 and that the Toda Hamiltonian $H$ (\ref{1.1}) satisfies $H\circ \cR = \iota_{\hat S}^*(\cH_2)$
 with (\ref{HR}), which are immediate consequences of the reduction.

To make the above formula explicit, we need to calculate
the minors $m_k(\Gamma^t\Gamma)$.
For this,  it is convenient to write
\be
X(\hat p, \hat q) = Y(\hat p, w(\hat p, \hat q))
\quad\hbox{with}\quad
Y(\hat p, w)=
\Gamma(\hat p, w)^t \Gamma(\hat p, w) =  V(\hat p)^t W W V(\hat p),
\label{YX}\ee
using the matrices  $W$ and $V$ defined in (\ref{VW}).
Since $Y_k$ is  the product of the $k\times n$ matrix obtained
by deleting the
last $(n-k)$ rows of $V^t W$ and its transpose, we can effortlessly calculate
$m_k(Y)=\det(Y_k)$ by applying
the standard Cauchy-Binet formula (e.g.~\cite{Gant}).
Letting  $I = \{ i_1,\ldots, i_k\}$,
$1 \leq i_1 <  \cdots <i_k \leq n$, run over the subsets
of $\{1,\ldots, n\}$,
the Cauchy-Binet formula now reads
\be
\det(Y_k) = \sum_{\vert I\vert =k} [\det(W_I) \det(V_I)]^2,
\ee
where $W_I$ is the $k\times k$ diagonal matrix with entries $w_{i_a}$ and  $V_I$
is the $k\times k$ Vandermonde matrix with entries
  $(V_I)_{a,b} = \hat p_{i_a}^{b-1}$.
Thus we  find the following result:
\be
m_k(Y(\hat p, w))=
\sum_{\vert I\vert =k}\biggl(\prod_{l\in I}w_l^2 \prod_{\substack{i,j\in I\\i\neq j}}\vert
\hat p_i-\hat p_j\vert\biggr).
\label{KFF}\ee
Substitution of  $w_l(\hat p,\hat q)$ from (\ref{wpar}), then directly leads to the
formula
\be
m_k(X(\hat p, \hat q)) =
\sum_{\vert I \vert =k} e^{\sum_{l\in I} \hat q_l} \!\! \prod_{i\in I, j\notin I}
\vert \hat p_i - \hat p_j\vert^{-1},
\ee
which is just the expression $\sigma_k$ in (\ref{1.4}).

Finally, the comparison of equations (\ref{1.3}), (\ref{1.4}) and (\ref{qkcircR}), (\ref{pcircR})
shows
that the gauge transformation $\cR \colon \hat S \to S$
is  nothing but the action-angle map $R\colon\hat M \to M$ derived
by Ruijsenaars in \cite{SR-90} with the aid of a different method.

\section{Group-theoretic interpretation of Toda duality}

We have seen that the Hamiltonian reduction approach to the open Toda lattice
permits an alternative derivation of the  explicit action-angle map (\ref{1.3}).
As a useful spin-off,  we can now interpret the
Toda duality in the same manner as was done previously for
Calogero type systems.

Remember that we have two natural Abelian Poisson algebras, $\{ \cH_k\}$ in (\ref{cH2}) and
$\{\hat \cH_k\}$ in (\ref{hatcHk}),
 before reduction.  Their generators are $N_+ \times K$ invariant
functions possessing complete Hamiltonian flows.
These Abelian algebras survive the reduction and
any generator of them descends to a Liouville integrable reduced Hamiltonian.
The only statement to check is that the generators of the reduced Abelian Poisson algebras
consist of $n$ independent functions, which  is readily verified using
the two alternative models  of $(T^*G)_\red$ given by
the Toda gauge and the Moser gauge (\ref{ident}).
As  detailed below,
these two Abelian algebras on $(T^*G)_\red$ engender a dual pair of integrable systems.

In terms of the model
$(S, \iota_S^*(\Omega))\equiv (M,\omega)$ of  $((T^*G)_\red, \Omega_\red)$,
the reduction of the Abelian Poisson algebra $\{\cH_k\}$
yields the commuting Toda Hamiltonians and the reduction
of $\{ \hat \cH_k\}$ becomes equivalent to the algebra of the position-variables
of the Toda system (\ref{1.1}).
In terms of the alternative model
$(\hat S, \iota_{\hat S}^*(\Omega))\equiv (\hat M,\hat \omega)$ of  $((T^*G)_\red, \Omega_\red)$,
the reduction of $\{\hat \cH_k\}$ gives the commuting Hamiltonians of the dual many-body
system (\ref{1.5}), whose position-variables are equivalent to the reduction of the algebra $\{ \cH_k\}$.
This means that the components of $q$ can be viewed both as position-variables
 of the Toda system  and as action-variables of the dual system,
 whose main Hamiltonian $\sigma_1$ (\ref{1.5}) is the reduction of $\hat \cH_1$.
Similarly,
the components of $\hat p$ encode action-variables for the Toda system and position-variables
for the dual system.
Thus one has two integrable many-body systems living on each other's action-angle phase spaces.
This is the essence of the duality relation discovered originally by Ruijsenaars
\cite{SR-90,SR-88}.

The commuting Hamiltonians of the systems in duality are generated by the Jacobi matrix
$L(q,p)$ (\ref{Lqp}) and the Hankel matrix
$X(\hat p, \hat q)= Y(\hat p, w(\hat p, \hat q))$ (\ref{YX}).
Here, it should be noted  that the  matrix $Y$, which has the  entries
$Y_{i,j} =\sum_{k=1}^n  (\hat p_k)^{i+j-2} w_k^2$,
appears in many papers dealing with Toda systems and related questions.  Indeed, equation
(\ref{mk}) represents the key ingredent of the reconstruction of the Jacobi matrix from its spectral data, which
goes back to classical work by Stieltjes (see e.g.~\cite{BSS} and references therein).
One could construct the Toda action-angle map by relying  directly on the relations (\ref{2.32}) and (\ref{mk}).
In our work we obtained these relations  by following the
standard procedure of Hamiltonian reduction.

We conclude with a list of open problems.
First, it could be interesting to apply quantum Hamiltonian reduction
\cite{STS} to gain a better understanding
of the quantum mechanical version of Toda duality.
For the state of art of this subject,  see the papers \cite{Bab-LMP,HR,Skly} and references therein.
Second, it would be important to generalize the group-theoretic
framework
presented in this paper so as to accommodate the open relativistic Toda lattice, whose dual
was also derived in \cite{SR-90}.
Finally, we remark that action-angle duals of closed Toda
lattices are not yet known, and this issue, as well as the cases of other root systems,
should be investigated in the future.
For constructions of action-angle variables of closed Toda lattices,  the reader may consult,
e.g.,~\cite{BBT, HK}.

\smallskip
\bigskip
\noindent{\bf Acknowledgements.}
I am indebted to C. Klim\v c\'\i k for important technical clarifications.
I thank the referees for useful comments.
I also wish to thank  I. Tsutsui for discussions and for suggesting
the term ``Moser gauge''.
This work was supported in part by the Hungarian Scientific
  Research Fund (OTKA) under the grant K 77400 and by the
  project TAMOP-4.2.2.A-11/1/KONYV-2012-0060 financed by the EU
  and co-financed by the European Social Fund.

\renewcommand{\theequation}{\arabic{section}.\arabic{equation}}
\renewcommand{\thesection}{\Alph{section}}
\setcounter{section}{0}

\section{Calculation of the symplectic form in the Moser gauge}
\setcounter{equation}{0}
\renewcommand{\theequation}{A.\arabic{equation}}

In this appendix we present the derivation of the formula (\ref{symp1}).
We start by noting that
direct substitution of $\cJ=\Lambda(\hat p)$ and $g^{-1}=\Gamma(w,\hat p)= W V(\hat p)$
into the symplectic form
$\Omega$ (\ref{2.5}) yields
\be
\iota_{\hat S}^* (\Omega) = - 2 d \tr (\Lambda dW W^{-1}) - 2 d \tr (\Lambda d V V^{-1}).
\label{A1}\ee
The first term is trivial to evaluate. The second term can be spelled out as
\be
\tr (\Lambda d V V^{-1}) = \sum_{i,j=1}^n \hat p_i (dV_{ij}) \check{V}_{ij} (\det V)^{-1}
\label{A2}\ee
where $\check{V}_{ij}$ the co-factor entering the inverse matrix,
$(V^{-1})_{ji} = \frac{\check V_{ij}}{\det V}$.
Since (on account of (\ref{VW}))  $V_{ij}$ depends only on $\hat p_i$ and
thus $\check{V}_{ij}$
does not depend on $\hat p_i$,
the  expression (\ref{A2}) can be recast as
\be
\tr (\Lambda d V V^{-1}) =
\sum_{i=1}^n   \hat p_i  (d \hat p_i)
\bigl(\partial_{\hat p_i} \bigl(\sum_{j=1}^n V_{ij} \check{V}_{ij}\bigr)\bigr)
 (\det V)^{-1}   =
\sum_{i=1}^n \hat p_i  (d \hat p_i) \partial_{\hat p_i} \ln \vert \det V \vert.
\label{A3}\ee
Therefore
\be
d \tr(\Lambda dV V^{-1}) =
\sum^n_{\substack{i,j=1\\i\neq j}} \hat p_i \left( \partial_{\hat p_j} \partial_{\hat p_i}
\ln \vert \det V \vert \right) d \hat p_j \wedge d \hat p_i
= \frac{1}{2}\sum^n_{\substack{i,j=1\\i\neq j}} \frac {d \hat p_j \wedge d \hat p_i}{\hat p_i - \hat p_j}.
\label{A4}\ee
The last equality was obtained using
 the Vandermonde determinant
 $\det V  = \prod_{ a < b}(\hat p_b - \hat p_a)$.

{}


\begin{thebibliography}{}

\bibitem{BBT}
O. Babelon, D. Bernard and M. Talon,
Introduction to Classical Integrable Systems,
Cambridge University Press, 2003

\bibitem{SR-90}
S.N.M.~Ruijsenaars,
Commun. Math. Phys. {\bf 133} (1990)  217


\bibitem{Mos}
J. Moser,  Lect. Notes in Phys. {\bf 38} (1975) 467


\bibitem{Kos}
B. Kostant, Adv. Math. {\bf 34} (1979) 195

\bibitem{OP}
M.A. Olshanetsky and A.M. Perelomov,
Invent. math. {\bf 54} (1979) 261

\bibitem{Bog}
O.I. Bogoyavlensky,
Commun. Math. Phys. {\bf 51} (1976) 201

\bibitem{Per}
A.M. Perelomov,
Integrable Systems of Classical Mechanics and Lie Algebras,
Birkh\"auser, 1990

\bibitem{DLT}
P. Deift, L.-C. Li and C. Tomei,
pp.~511-536 in: Important Developments in Soliton Theory,
A.S. Fokas and V.E. Zakharov (Eds.), Springer, 1993

\bibitem{RSTS}
A.G. Reyman and  M.A. Semenov-Tian-Shansky, pp.~116-259 in:
Encyclopedia of Mathematical Sciences,
Vol. 16, V.I. Arnold and S.P. Novikov (Eds.),  Springer, 1994


\bibitem{JHEP}
V.~Fock, A.~Gorsky, N.~Nekrasov and V.~Rubtsov,
JHEP {\bf 0007} (2000)  028


\bibitem{FK-NPB}
L.~Feh\'er and C.~Klim\v c\'\i k,
Nucl. Phys. B {\bf 860} (2012) 464

\bibitem{SR-88}
S.N.M.~Ruijsenaars,
Commun. Math. Phys. {\bf 115} (1988)  127

\bibitem{RIMS95}
S.N.M.~Ruijsenaars,
Publ. RIMS  {\bf 31} (1995) 247


\bibitem{DLNT}
P. Deift, L.-C. Li, T. Nanda and C. Tomei,
Comm. Pure Appl. Math. {\bf XXXIX} (1986) 183


\bibitem{FG}
L. Faybusovich  and M. Gekhtman,
 Phys. Lett. A {\bf 272}  (2000) 236



\bibitem{Tsig}
Y.A. Grigoryev and A.V. Tsiganov,
SIGMA {\bf 2} (2006) 097

\bibitem{Skly}
E. Sklyanin, 
J. Phys. A: Math. Theor. {\bf 46} (2013) 382001


\bibitem{Nanda}
T. Nanda, Siam. J. Numer. Anal. {\bf 22} (1985) 310

\bibitem{Gant}
F.R. Gantmacher, The Theory of Matrices, Vol 1, Chelsea Publ. Co., 1959

\bibitem{BSS}
R. Beals, D.H. Sattinger and J. Szmigielski,
Comm. Pure Appl. Math. {\bf LIV} (2001) 0091


\bibitem{STS}
M.A. Semenov-Tian-Shansky,
pp. 226-259 in:
 Encyclopedia of Mathematical Sciences,
Vol. 16, V.I. Arnold and
S.P. Novikov (Eds.),  Springer, 1994

\bibitem{Bab-LMP}
O. Babelon, Lett. Math. Phys. {\bf 65} (2003) 229

\bibitem{HR}
M. Halln\"as and S. Ruijsenaars,
Journ. Math. Phys. {\bf 53} (2012) 123512

\bibitem{HK}
A. Henrici  and T. Kappeler,
Int.~Math.~Res.~Not. {\bf Vol. 2008} (2008) article ID rnn031


\end{thebibliography}
\end{document}